\documentclass{article}
\usepackage{authblk}
\usepackage[utf8]{inputenc}
\usepackage[numbers,super]{natbib}
\usepackage{amssymb}
\usepackage{amsmath}
\usepackage{graphicx}
\usepackage{mdwlist}
\usepackage{tabularx}
\usepackage{caption}
\usepackage{subcaption}
\usepackage{wasysym}
\usepackage[switch]{lineno}

\begin{document}

\title{Temporally distributed parent body accretion in the C reservoir of the solar system}
\author[1,2,3,*]{Wladimir Neumann}
\author[4]{Ning Ma}
\author[5]{Audrey Bouvier}
\author[2]{Mario Trieloff}
\affil[1]{Institute of Geodesy and Geoinformation Science, Technische Universität Berlin, Kaiserin-Augusta-Allee 104-106, 10553 Berlin, Germany}
\affil[2]{Klaus-Tschira-Labor für Kosmochemie, Institut für Geowissenschaften, Universität Heidelberg, Im Neuenheimer Feld 234-236, 69120 Heidelberg, Germany}
\affil[3]{Institute of Planetary Research, German Aerospace Center (DLR), Rutherfordstr. 2, 12489 Berlin, Germany}
\affil[4]{Institute of Geochemistry and Petrology, ETH Zürich, Sonneggstrasse 5, 8092 Zürich, Switzerland}
\affil[5]{Bayerisches Geoinstitut, University of Bayreuth, 95440 Bayreuth, Germany}
\affil[*]{Corresponding author, wladimir.neumann@dlr.de, orcid=0000-0003-1932-602X, www.researchgate.net/profile/Wladimir\_Neumann}
\date{}
\setcounter{Maxaffil}{0}
\renewcommand\Affilfont{\itshape\small}

\maketitle

\begin{abstract}
Accretion processes in protoplanetary disks produce a diversity of small bodies that played a crucial role in potentially multiple reshuffling events throughout the solar system and in both early and late accretion of planets. Application of thermo-chronometers to meteorites provides precise dating of the formation age of various mineralogical components. Nucleosynthetic isotopic anomalies that indicate a dichotomy between non-carbonaceous (NC) and carbonaceous (C) meteorites and precise parent body chronology can be combined with planetesimal thermal evolution models to constrain the timescale of accretion and dynamical processes in the early solar system.
Achondrite parent bodies are considered to have accreted early and mostly in the NC region. By contrast, late accretion in the C region produced mostly undifferentiated parent bodies, such as the parent body of CR chondrites that formed as late as 4 Ma after solar system formation\cite{Schrader2011}. However, presence of more evolved CR-related meteorites suggests also an earlier accretion timing.
We present modeling evidence for a temporally distributed accretion of parent bodies of CR-related meteorite groups that originate from a C reservoir and range from aqueously altered chondrites to highly equilibrated chondrites and partially differentiated primitive achondrites. The parent body formation times derived range from $<1$ Ma to $\approx 4$ Ma after solar system formation, with $\approx 3.7$ Ma, $\approx 1.5-2.75$ Ma, $\lesssim 0.6$ Ma, and $\lesssim 0.7$ Ma for CR1-3, Flensburg, NWA 6704, and NWA 011. This implies that accretion processes 
in the C reservoir started as early as in the NC reservoir and produced differentiated parent bodies with carbonaceous compositions in addition to undifferentiated C chondrite parent bodies. The accretion times correlate inversely with the degree of the meteorites' alteration, metamorphism, or differentiation. The accretion times for the CI/CM, Ryugu, and Tafassites parent bodies of $\approx 3.75$ Ma, $\approx 1-3$ Ma, and $1.1$ Ma\cite{Ma2022,Neumann2021}, respectively, fit well into this correlation in agreement with the thermal and alteration conditions suggested by the meteorites.
\end{abstract}

\section{One-Sentence Summary} Thermal evolution fits to the chronology of CR-related meteorites reveal accretion times of $<1$ to $4$ Ma in the outer solar system.

\section{Highlights}
\begin{itemize}
\item We fit thermal evolution models to the thermo-chronological data of CR-related meteorites CR1-3, NWA 011, NWA 6704, and Flensburg
\item Our modeling suggests different parent bodies with different degrees of alteration, metamorphism, or differentiation
\item Parent body accretion times derived range from $<1$ Ma to $\approx 4$ Ma after solar system formation, with $\approx 3.7$ Ma, $\approx 1.5-2.75$ Ma, $\lesssim 0.7$ Ma, and $\lesssim 0.6$ Ma for CR1-3, Flensburg, NWA 6704, and NWA 011, respectively
\item Accretion processes in the C reservoir started as early as in the NC reservoir and produced differentiated parent bodies with carbonaceous compositions
\item Accretion times of carbonaceous chondritic CI/CM, Ryugu, and Tafassites parent bodies fit well into the sequence derived
\end{itemize}

Keywords: Asteroids; Meteorites; Chronology; Accretion; Parent Bodies.

\begin{figure}
\begin{minipage}[ht]{6.5cm}
\setlength{\fboxsep}{0mm}
\includegraphics[trim = 11.6mm 0mm 3.5mm 0mm, clip, width=6.5cm]{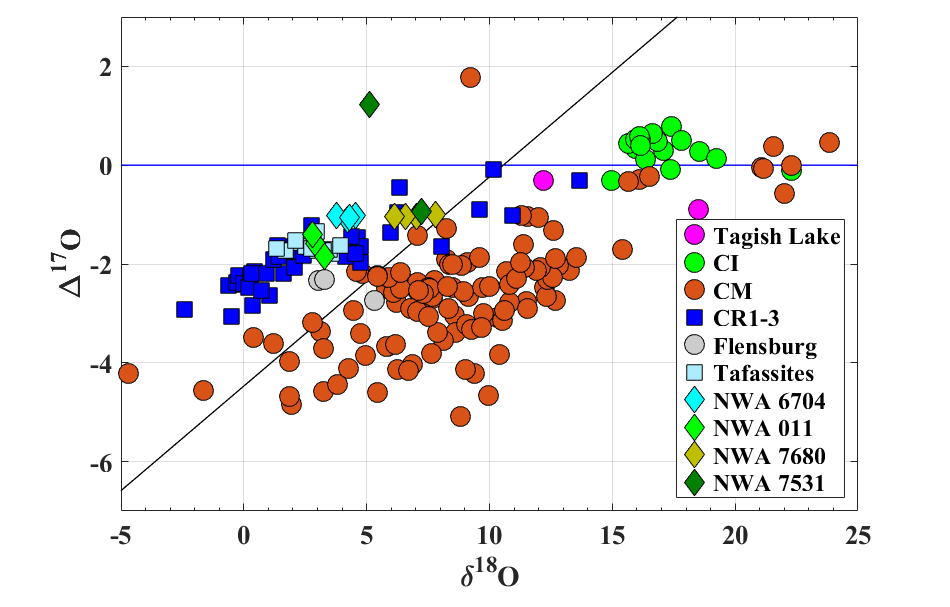}
\end{minipage}
\begin{minipage}[ht]{6.5cm}
\setlength{\fboxsep}{0mm}
\centerline{\includegraphics[trim = 10.6mm 0mm 5mm 0mm, clip, width=6.5cm]{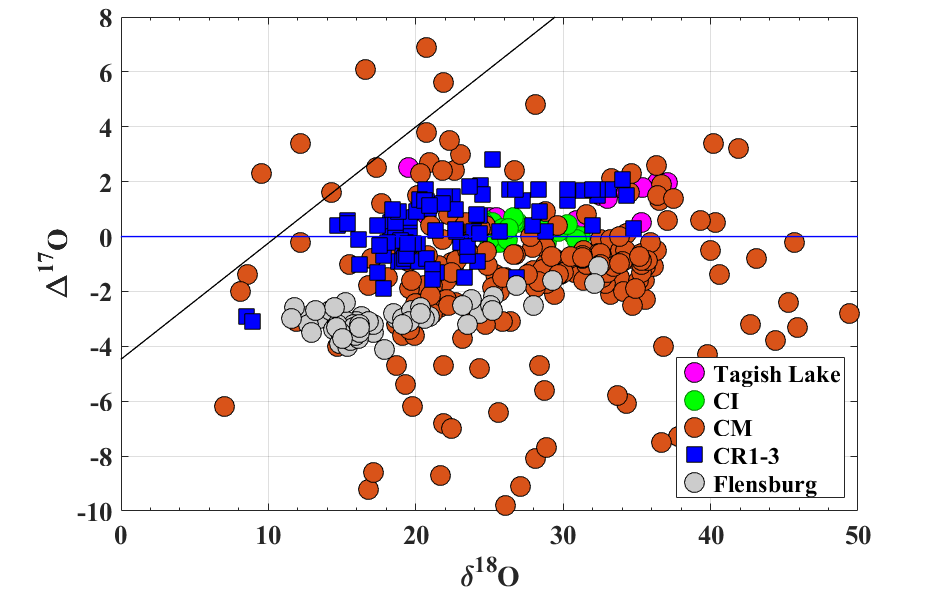}}
\end{minipage}
\\
\begin{minipage}[ht]{6.7cm}
\setlength{\fboxsep}{0mm}
\centerline{\includegraphics[trim = 12.5mm 5mm 0mm 5mm, clip, width=6.6cm]{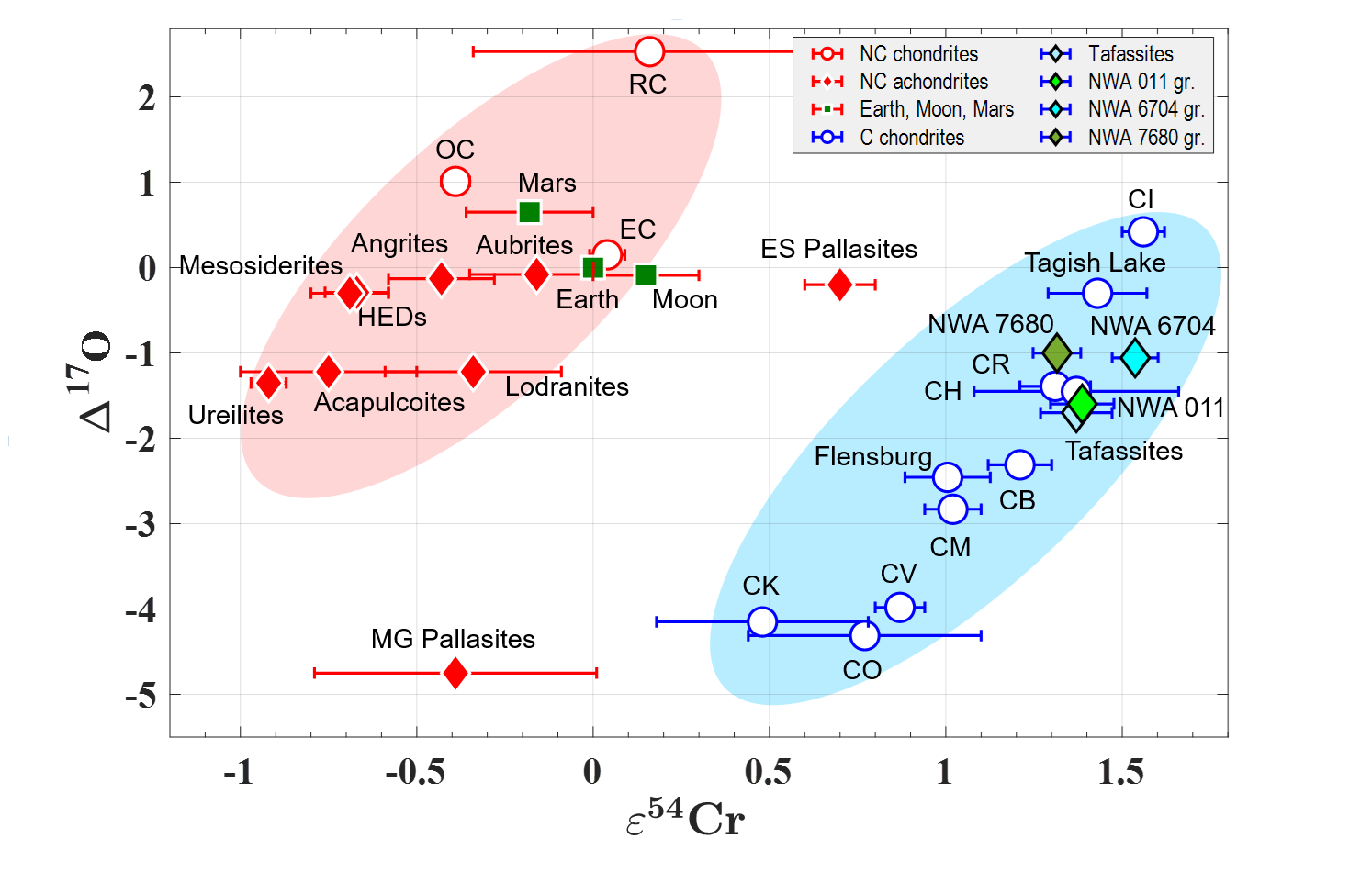}}
\end{minipage}
\caption{Bulk (\textit{top left}) and carbonate (\textit{top right}) oxygen isotope data of CR-related meteorites in comparison with CM, CI, Tagish Lake, and Flensburg. \textit{Bottom}: Comparison of the $\Delta^{17}$O-$\varepsilon^{54}$Cr systematics of CR-related carbonaceous achondrites with other C and NC materials. The oxygen isotope data were obtained from the Meteoritical Bulletin database, access September 2022, and \cite{Rowe1994,Clayton1999,Bischoff2021,Tyra2007,Tyra2012,Tyra2013,Tyra2016,Schrader2011,Baker2002,Hyde2013,Engrand2001,Leshin2001,Yurimoto2008,Piralla2020,Telus2019,Lindgren2017,Fujiya2019,Benedix2003,Jilly2018,Ma2022}. The CR-related meteorite bulk O isotopic compositions plot to the left of the carbonaceous chondrite anhydrous mineral lines (CCAM). The Flensburg data plot around CCAM overlapping marginally with the CR-related data, while TL, CI, and CM plot to the right of CCAM. All groups remain below the terrestrial fractionation line (TFL), except CI chondrites that plot slightly above it. The carbonates in CR1-3 have comparable $\delta^{18}$O values to Flensburg, scatter to both higher and lower values than CI, and to lower values than TL. CR1-3 carbonate data plot around TFL, similar to CI, but with a stronger scatter up and downwards. However, Flensburg carbonates plot far below the TFL, while the CM scatter encompasses every group. For $\Delta^{17}$O-$\varepsilon^{54}$Cr systematics, average values for groups or grouplets are plotted. The data are from \cite{Dauphas2016,Scott2018,Burkhardt2019,Sanborn2019,Huyskens2019}.
The meteorite Flensburg and CR-related groups Tafassites and CR, as well as grouplets NWA 011, NWA 6704, and NWA 7680 have C-like nucleosynthetic anomalies. The CR-related meteorites plot closely to each other and to CH chondrites. They all are within the same region of the C meteorite composition. The similar O isotopic composition and nucleosynthetic anomalies of CR-related meteorites reflect a similar accreting material within a relatively confined region further out in the protoplanetary disk, and, more broadly, a similar material with other C meteorites.}
\label{fig1}
\end{figure}

\paragraph{Introduction}
Meteoritic components provide evidence for the isotopical heterogeneity of the early solar nebula and carry $^{16}$O excesses derived from nucleosynthesis or CO self shielding.\cite{Clayton1999} Isotopic variations from inter-reservoir exchange give rise to linear mixing lines for different meteorite groups or clans in the oxygen three-isotope plot.\cite{Clayton1993} 
The CR chondrites is a group of undifferentiated meteorites that includes two subgroups defined based on their degree of aqueous alteration (CR1-3) or thermal metamorphism and partial melting (initially classified as CR6-7). Their common origin indicated by oxygen isotope systematics (Fig. \ref{fig1}, top panels) is challenged by a striking contrast in the metamorphism and alteration degree that could indicate distinct evolution paths and parent bodies. The CR1-3 bulk oxygen isotope compositions form a unique mixing line\cite{Clayton1999} that requires an anhydrous precursor different from CI and CM.\cite{Weisberg1993,Schrader2011} An oxygen isotopic heterogeneity\cite{Weisberg1995} and patterns similar to CM imply low-temperature aqueous alteration on an isotopically heterogeneous chondritic parent body (PB).\cite{Clayton1999} The CR6-7 data are broadly similar to the CR1-3 mixing line, but give rise to a separate mixing line and were suggested to belong to the carbonaceous primitive achondrite group called Tafassites.\cite{Ma2022}
Further meteorites with geochemical affinities to CR include the basaltic achondrite grouplets NWA 011,\cite{Weiss2013} NWA 6704,\cite{Sanborn2019} and NWA 7680,\cite{Huyskens2019} other members of the Tafassites group,\cite{Ma2022} and, potentially, the chondrite Flensburg.\cite{Bischoff2021} Considered together with CR1-3, these meteorites are clearly distinct from other carbonaceous meteorites. Precursors of NWA 011, NWA 6704 and NWA 7680 grouplets are related to CR based on either indistinguishable or broadly similar oxygen isotope compositions.
NWA 011, NWA 6704, and NWA 7680 are, further, related to CR based on the $\varepsilon ^{54}Cr$ similarity\cite{Huyskens2019} that is close to the ordinary and enstatite chondrites for all other differentiated meteorites, but not for these grouplets (and Tafassites\cite{Ma2022}). 
However, NWA 011 and its pairings likely do not originate from the same parent body as Tafassites, due to different FeO-Mn compositions and incompatible chemical composition, e.g., higher silicate FeO content and Ca-rich pyroxene unlikely to have formed from the same precursor. 
NWA 6704 grouplet’s $\Delta^{17}$O values suggest another separate parent body, supported, further, by its unique pyroxenitic lithology.
The Flensburg bulk data plots in the transition field to CR chondrites with a marginal overlap.\cite{Bischoff2021}

Therefore, the broad oxygen isotope composition similarity rather indicates that Tafassites, NWA 6704, NWA 011, NWA 7680, Flensburg, and CR chondrites derive from at least six different parent bodies. Here, we address four of those, while the Tafassites parent body was considered in \cite{Ma2022} and not enough data is available for NWA 7680.

The genetic relationship of CR-related meteorites implies that they sourced from the same region and their oxygen systematics were produced by an isotopic exchange between $^{16}$O-rich solids and $^{16}$O-poor gases\cite{Clayton1993}. In general, the bulk O isotope compositions of the CR-related meteorites plot in a more $^{16}$O-rich region relative to water-rich carbonaceous CI, CM, Tagish Lake, and Flensburg meteorites (Fig. \ref{fig1}, top left panel). Similar $\delta^{18}$O and $\delta^{17}$O values imply that all CR-related groups are derived from a single homogeneous region of the solar nebula. This region contained more $^{16}$O-rich material than the Earth and CI chondrites based on the position below TFL. Lower $\delta^{18}$O values suggest, further, lower condensation or equilibration temperature for the precursor material than for that of CI and CM chondrites. Formation in the C reservoir beyond the orbit of Jupiter has been suggested for CR-related meteorites.\cite{Bottke2017}

The carbonate O isotopic data indicates partially similar carbonate formation conditions concerning temperature and oxygen fugacity and carbon source in CR1-3 chondrites as in CI, CM, and TL chondrites, but different to Flensburg (Fig. \ref{fig1}, top right panel). Both implications agree with roughly similar alteration temperatures for CR1-3, CI, CM, and TL, but a higher alteration temperature of Flensburg. The differences in $\Delta^{17}$O and $\delta^{18}$O values between the groups imply different alteration conditions, e.g., the duration, temperature, redox conditions, and fluid composition experienced.\cite{Schrader2011}

Meteorite nucleosynthetic anomalies distinguish two major groups of meteorites, NC and C (Fig. \ref{fig1}, bottom panel), and two major reservoirs of planetesimal formation separated early on in the protoplanetary disk. The meteorite Flensburg and CR-related groups and grouplets (CR, Tafassites, NWA 011, NWA 6704, and NWA 7680) have C-like nucleosynthetic anomalies. The CR-related meteorites plot closely to each other and to CH chondrites. All these meteorites are within the same region of the C meteorite composition. Their similar O isotopic composition and nucleosynthetic anomalies reflect a similar accreting material within a relatively confined region further out in the protoplanetary disk. More broadly, they reflect a similar material with other C meteorites, such as Flensburg, CI, CM, and with C-type NEAs like Ryugu and Bennu.

\begin{table}
\centering
\caption{Time (in Ma rel. to CAIs) and temperature (in K) data used for fitting the CR-related meteorite parent bodies. The data for water-rich CI and CM chondrites and for CR-related Tafassites are shown as well. Notes: $^{(\text{b})}$ breunnerite, $^{(\text{c})}$ calcite, $^{(\text{d})}$ dolomite, $^{(\text{p})}$ phosphate, $^{(\text{pl})}$ plagioclase, $^{(\text{px})}$ pyroxene, $^{(\text{w})}$ whole-rock, $^{(\text{ol})}$ olivine, $^{(\text{ca})}$ carbonates. The age data are from \cite{Jilly2017}(CR1-3), \cite{Bischoff2021}(Flensburg), \cite{Bouvier2011}(NWA 011 and NWA 2976), \cite{Huyskens2019,Amelin2019a}(NWA 4587), \cite{Sanborn2019,Amelin2019}(NWA 6704), \cite{Huyskens2019}(NWA 10132), \cite{Bouvier2011,Ma2022}(Tafassites), \cite{Jilly2017,Neumann2021}(CI, CM). The closure temperatures are from \cite{Jilly2015,Jilly2018}(CR1-3), \cite{Bischoff2021}(Flensburg), \cite{Goepel1994,Dodson1973} and \cite{Steffens2019}(NWA 011, NWA 2976, NWA 4587, NWA 6704, NWA 10132), \cite{Ma2022}(Tafassites), \cite{Neumann2021}(CI, CM). A CAI age of 4567.9 Ma was used.}
\centering
\begin{tabularx}{\columnwidth}{l r r r r c}
\hline \\ [-1.5ex]
Meteorite & \multicolumn{2}{r}{Closure $T$} & \multicolumn{2}{r}{Closure $t$} & Method
\\
 & $T^{c}$ & $\sigma_{T}$ & $t^{c}$ & $\sigma_{t}$ \\
 & K & K & Ma & Ma \\ [-1.5ex]
\\ [-1.0ex]
\hline
\\ [-2.0ex]
\multicolumn{6}{c}{Aqueously altered CR1-3} \\
\\ [-2.5ex]
Renazzo $^{(\text{d})}$ & 344.5 & 16.5 & 4.8 & +4.24/-2.4 & Mn-Cr\\
Renazzo $^{(\text{c})}$ & 344.5 & 16.5 & 4.55 & +7.4/-2.8 & Mn-Cr\\
GRO 95577 $^{(\text{c})}$ & 308 & 25 & 12.57 & +2.1/-1.4 & Mn-Cr\\
\\ [-2.0ex]
\multicolumn{6}{c}{Aqueously altered CI and CM} \\
\\ [-2.5ex]
Orgueil, Ivuna, Y980115 $^{(d)}$ & 398 & 50 & 5.485 & 0.655 & Mn-Cr \\
Orgueil $^{(\text{b})}$ & 398 & 50 & 9.7 & 3.6 & Mn-Cr \\
Murchison, Y791198 $^{(c)}$ & 318 & 50 & 5.445 & 0.655 & Mn-Cr \\
\\ [-2.0ex]
\multicolumn{6}{c}{Aqueously altered Flensburg} \\
\\ [-2.5ex]
Flensburg $^{(\text{ca})}$ & 423 & 50 & 3.4 & 1 & Mn-Cr \\
\\ [-2.0ex]
\multicolumn{6}{c}{Partially molten Tafassites (T6-7)} \\
\\ [-2.5ex]
Tafassasset $^{(\text{w})}$ & 1200 & 100 & 2.9 & 0.9 & Hf-W \\
Tafassasset $^{(\text{w})}$ & 950 & 100 & 4.9 & 0.3 & Mn-Cr \\
Tafassasset $^{(\text{p})}$ & 720 & 50 & 19.86 & 8.1 & U/Pb-Pb \\
NWA 11561 $^{(\text{p})}$ & 720 & 50 & 7.55 & 2.85 & U/Pb-Pb \\
NWA 7317 $^{(\text{p})}$ & 720 & 50 & 8.42 & 2.94 & U/Pb-Pb \\
NWA 12455 $^{(\text{p})}$ & 720 & 50 & 10.1 & 1.75 & U/Pb-Pb \\
\\ [-2.0ex]
\multicolumn{6}{c}{Differentiated CR-related grouplets} \\
\\ [-2.5ex]
NWA 011 $^{(\text{pl, px})}$ & 750 & 130 & 4.5 & 0.33 & Al-Mg \\
NWA 011 $^{(\text{ol, px})}$ & 850 & 100 & 5.62 & 2.64 & Mn-Cr \\
NWA 2976 $^{(\text{w, pl, px})}$ & 750 & 130 & 5.08 & 0.05 & Al-Mg \\
NWA 2976 $^{(\text{pl, px})}$ & 770 & 50 & 5.01 & 0.59 & U/Pb-Pb \\
NWA 4587 $^{(\text{w, pl, px})}$ & 750 & 130 & 4.68 & 0.04 & Al-Mg \\
NWA 4587 $^{(\text{px})}$ & 770 & 50 & 5.47 & 0.23 & U/Pb-Pb \\
\\ [-1.5ex]
NWA 6704 $^{(\text{pl, px})}$ & 750 & 130 & 5.31 & 0.13 & Al-Mg \\
NWA 6704 $^{(\text{pl, px})}$ & 850 & 100 & 6.24 & 0.81 & Mn-Cr \\
NWA 6704 $^{(\text{px})}$ & 770 & 50 & 5.14 & +0.3/-0.22 & U/Pb-Pb \\
NWA 10132 $^{(\text{pl, px, ol})}$ & 750 & 130 & 5.42 & 0.7 & Al-Mg \\
\\ [-1.5ex]
NWA 7680 & 850 & 100 & 5.53 & 0.34 & Mn-Cr \\
\hline
\end{tabularx}
\label{table1}
\end{table}

\paragraph{Data and methods:} Characteristic parent body properties were derived from the analysis of formation of various primary (olivine, pyroxene, plagioclase) or secondary (carbonates, such as calcite, dolomite, or breunnerite, formed in the presence of aqueous solutions) mineralogical components, or whole-rock samples (Table \ref{table1}). Carbonates formed close to the peak of hydrothermal activity in the source regions of CR1-3 and Flensburg chondrites. Since they incorporate Mn, the short-lived decay system $^{53}$Mn-$^{53}$Cr with a $^{53}$Mn half-life of $3.7$ million years is used to constrain their age. Data available for CR give rise to a Renazzo dolomite, a GRO 95577 calcite, and a Renazzo calcite data points.
Calcite and dolomite formation in Flensburg has been dated with almost indistinguishable relative formation times that give rise to a single data point.\cite{Bischoff2021} In the heavily metamorphosed and partially molten T6-7 meteorites and differentiated CR-related NWA 011, NWA 6704, and NWA 7680 grouplets that lack carbonates, other components, such as whole-rock, olivine, pyroxene, plagioclase, and phosphates were dated with the U/Pb-Pb, Al-Mg, and Mn-Cr chronometers (Table \ref{table1}). U/Pb-Pb phosphate data were derived recently,\cite{Ma2022} providing one data point for each of three North-West Africa Tafassites and Tafassasset in addition to previously available Hf-W and Mn-Cr data, while for the differentiated CR-related grouplets Al-Mg, Mn-Cr, or Pb-Pb data are available. Overall formation temperatures for all these components are bracketed by $\approx 300$ K and $\approx 1300$ K, and the ages available fall into the first $\approx 27$ Ma after the solar system formation (Table \ref{table1}). Thus, the data describe the cooling behavior after reaching $^{26}$Al-induced temperature maxima at the burial depths of respective meteorites, except for Flensburg, where carbonates formed during the initial heating phase triggered by $^{26}$Al on the prograde branch of a temperature curve. \cite{Bischoff2021} The carbonate ages in CI and CM meteorites fit to the above time scale \cite{Jilly2017} providing a possibility for a comparison of parent body aqueous alteration history.\cite{Neumann2021}
\begin{figure}
\begin{minipage}[ht]{6cm}
\setlength{\fboxsep}{0mm}
\centerline{\includegraphics[trim = 10mm 0mm 10mm 0mm, width=5.6cm]{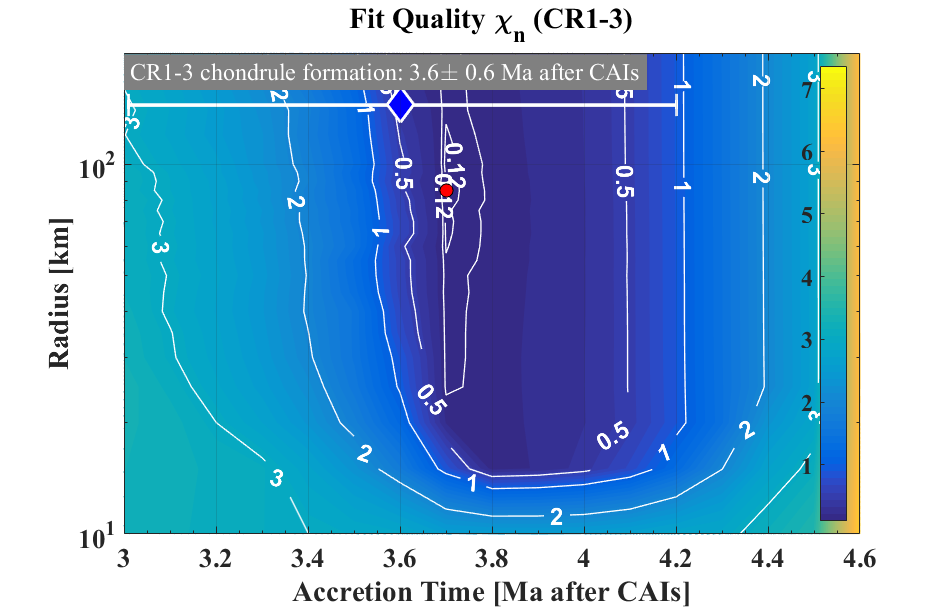}}
\end{minipage}
\begin{minipage}[ht]{6cm}
\setlength{\fboxsep}{0mm}
\centerline{\includegraphics[trim = 10mm 0mm 10mm 0mm, width=5.6cm]{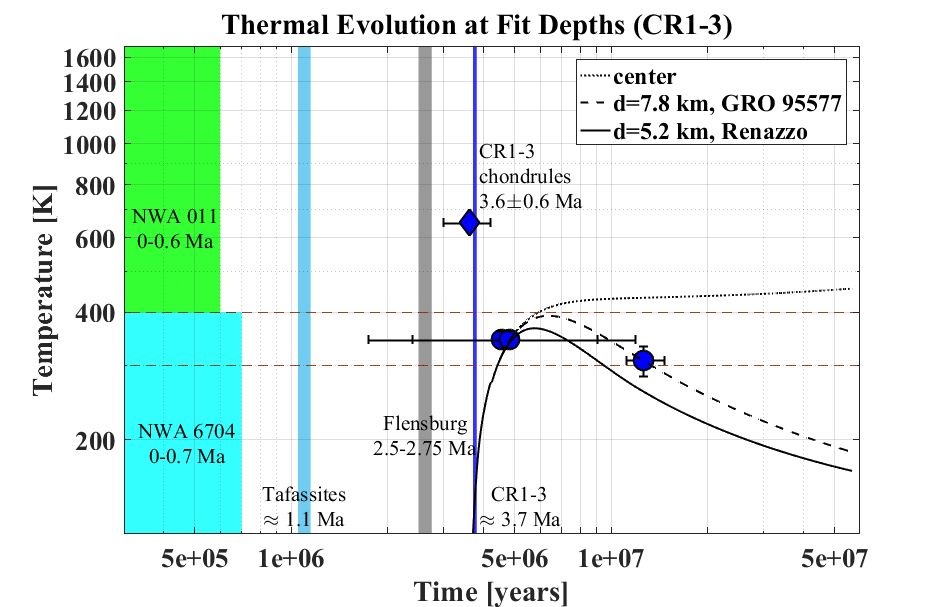}}
\end{minipage}
\\
\begin{minipage}[ht]{6cm}
\setlength{\fboxsep}{0mm}
\centerline{\includegraphics[trim = 10mm 0mm 10mm 0mm, width=5.6cm]{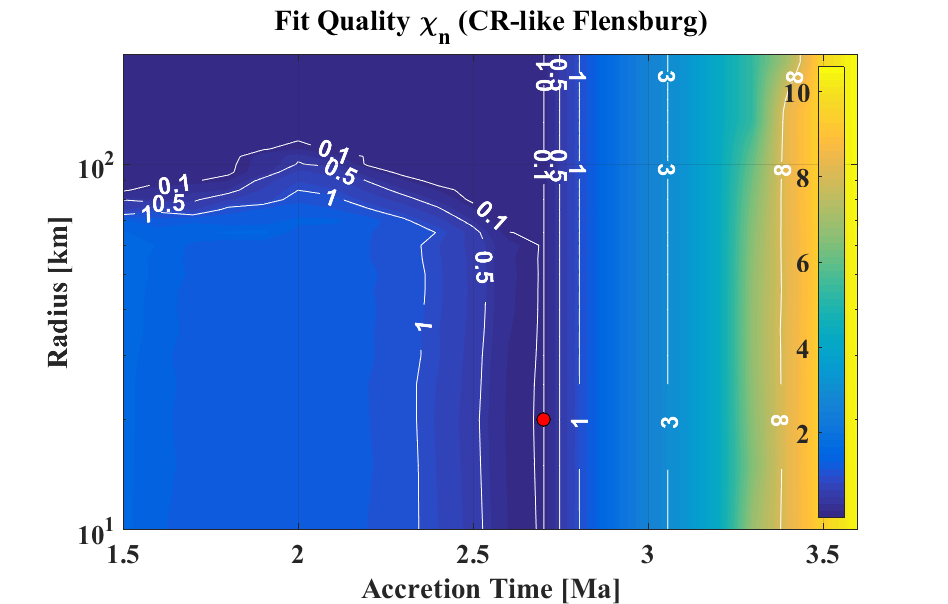}}
\end{minipage}
\begin{minipage}[ht]{6cm}
\setlength{\fboxsep}{0mm}
\centerline{\includegraphics[trim = 10mm 0mm 10mm 0mm, width=5.6cm]{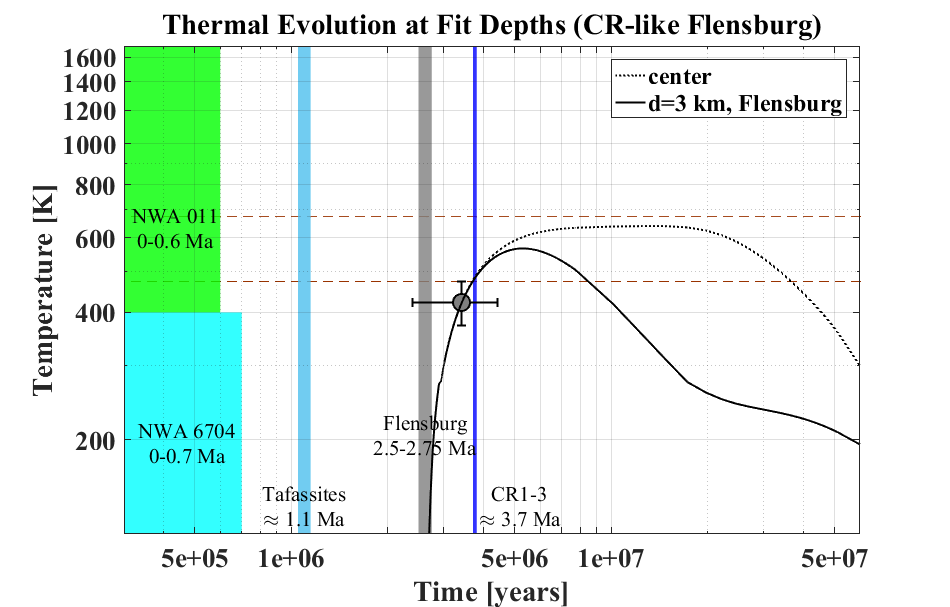}}
\end{minipage}
\\
\begin{minipage}[ht]{6cm}
\setlength{\fboxsep}{0mm}
\centerline{\includegraphics[trim = 10mm 0mm 10mm 0mm, width=5.6cm]{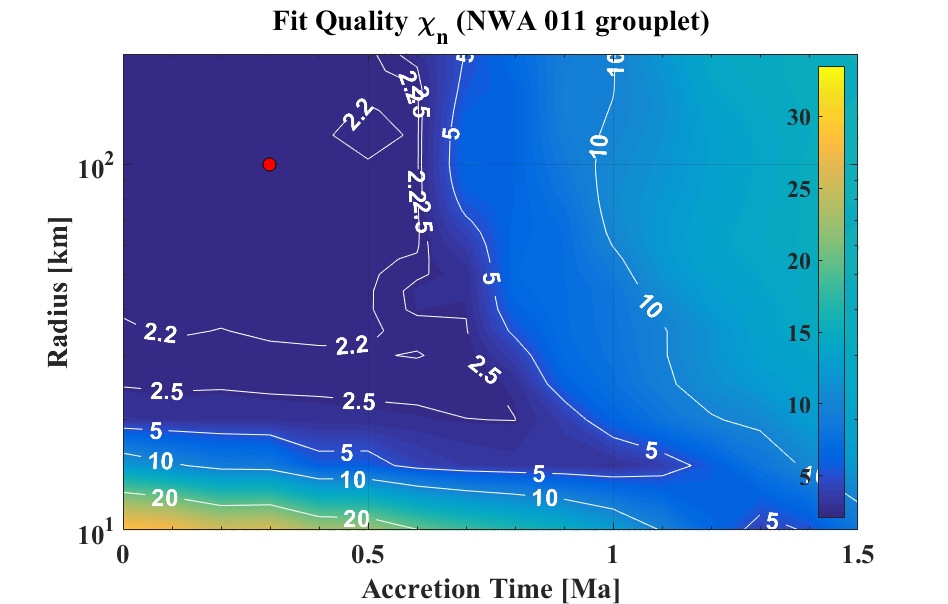}}
\end{minipage}
\begin{minipage}[ht]{6cm}
\setlength{\fboxsep}{0mm}
\centerline{\includegraphics[trim = 10mm 0mm 10mm 0mm, width=5.6cm]{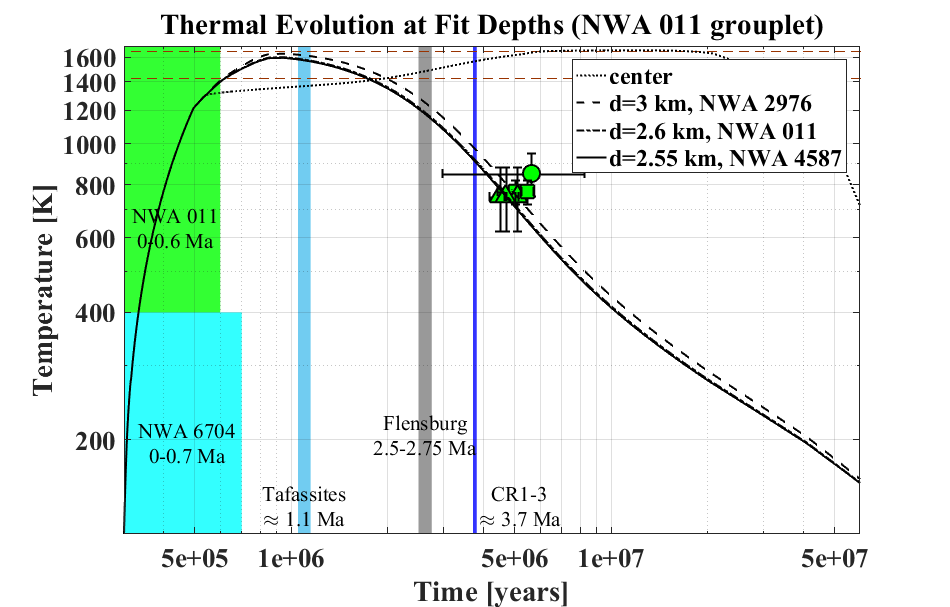}}
\end{minipage}
\\
\begin{minipage}[ht]{6cm}
\setlength{\fboxsep}{0mm}
\centerline{\includegraphics[trim = 10mm 0mm 10mm 0mm, width=5.6cm]{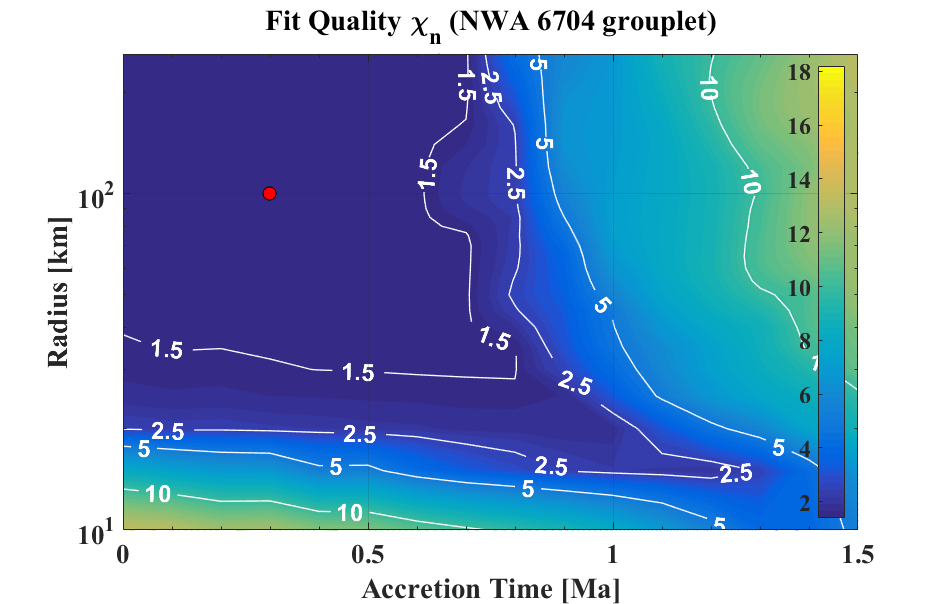}}
\end{minipage}
\begin{minipage}[ht]{6cm}
\setlength{\fboxsep}{0mm}
\centerline{\includegraphics[trim = 10mm 0mm 10mm 0mm, width=5.6cm]{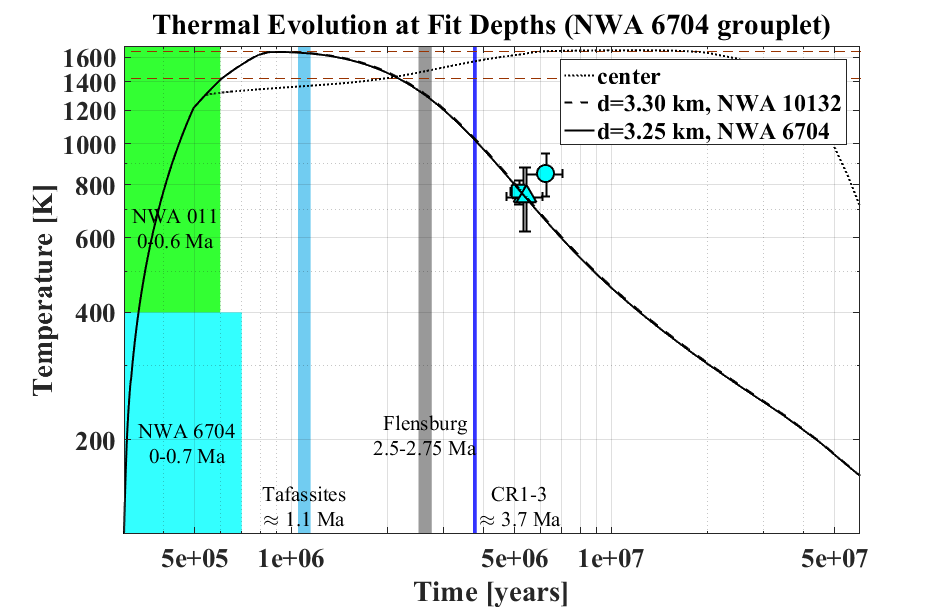}}
\end{minipage}
\caption{Left column: Both colorbar and isolines show fit quality as a function of planetesimal radius $R$ and accretion time $t_{0}$ for CR1-3 (first row),  Flensburg (second row), NWA 011 (third row), NWA 6704 (fourth row). Best-fit accretion times correspond from top to bottom to $\approx 3.7$ Ma, $\approx 2.5-2.75$ Ma, $\lesssim 0.6$ Ma, and $\lesssim 0.7$ Ma. The exemplary parent bodies from the respective minimum fit quality regions with $R=90$ km and $t_{0}=3.7$ Ma, $R=20$ km and $t_{0}=2.7$ Ma, $R=100$ km and $t_{0}=0.3$ Ma, as well as $R=100$ km and $t_{0}=0.3$ Ma, from top to bottom, are indicated with red dots. Right column: Thermal evolution at the fit depths $d$ of respective meteorites for best-fit parent bodies indicated by red dots in the left column. Color patches show PB accretion time intervals. The accretion time of the CR-related Tafassite PB is included for completeness.}
\label{fig2}
\end{figure}
We used state-of-the-art global thermal evolution models for early solar system planetesimals. The numerical setup is described in detail in the Supplementary Material. The models included, metal-silicate separation in the case of NWA 011 and NWA 6704, as melting and (partial) differentiation are indicated by the meteorites. Properties of CR-related parent bodies were obtained by approximating the data points with the evolution of the temperature at different depths in the respective time interval using a root mean square (RMS) procedure and by comparing with the maximum metamorphic temperatures.\cite{Henke2012,Neumann2018b,Gail2019} By varying the radius $R$ over a wide range of $10$ km to $200$ km and the accretion time $t_{0}$ between $0$ and $5$ Ma after CAIs, we constrain ranges within the $(R,t_{0})$-diagram appropriate for bodies that are likely to have produced various CR-related materials.

\paragraph{Results:}
The fit quality calculated (Fig. \ref{fig2}) shows best-fit parent body accretion times and sizes. The absolute value of the optimal fit quality is case dependent, while, in general, best-fit objects are characterized by its minimum. A global minimum is difficult to obtain given that the number of data points available varies between $1$ and $6$ for different parent bodies. Thus, a minimum or best-fit field defined by a plateau of the fit quality and an acceptable fit quality defined by strong gradients around the boundaries of the plateau are more suitable than a global minimum. Conclusive constraints on the accretion time were obtained with overall uncertainties of $\lesssim 1$ Ma for each meteorite group, while the radius estimates are less ideal. For CR1-3, a field with a minimum fit quality $\chi_{n}\leq 0.15$ is obtained for an accretion between $3.7$ and $3.8$ Ma after CAIs (Fig. \ref{fig2}, top left panel). The corresponding radius spread is $30-200$ km, but a minimum of $0.12$ for $\chi_{n}$ is obtained at $t_{0}\approx 3.7$ Ma and $R=70-120$ km. For an acceptable fit quality of $\leq 0.5$, the accretion time is bracketed by $3.6$ and $4.1$ Ma. Overall, the best-fit accretion time of $3.7$ Ma is almost indistinguishable from the CR1-3 chondrule formation time of $\approx 3.6$ Ma.\cite{Budde2018}

In the case of Flensburg, an acceptable $\chi_{n}\leq 0.5$ field is obtained between $t_{0}=2.5$ and $t_{0}=2.75$ Ma for any radius, but also for an accretion as early as $1.5$ Ma of bodies larger than $R=80$ km (Fig. \ref{fig2}, second row, left panel). Thus, no constraint on the radius can be derived from the fit procedure. The reason for that and for an acceptable fit quality for an early accretion is the availability of only one data point. For any object within the field with $\chi_{n}\leq 0.5$, a depth can be found at which the temperature curve crosses the center of the data point on the ascending branch. However, the earlier the accretion, the shallower this depth (down to $\lesssim 300$ m) and the higher the peak central temperature of the planetesimal, correlating inversely with the accretion time and exceeding the melting temperature in extreme cases. Thus, we consider planetesimals that accreted prior to $2.5$ Ma as less likely parent body candidates. Those that accreted between $2.5$ Ma and $2.75$ Ma have, by contrast, mostly homogeneous structures and thermal conditions throughout their interiors, in agreement with the prevalent notion of water-rich chondrite parent bodies. An accretion time of $\approx 2.7$ Ma fits the Flensburg carbonate data best for any parent body size.

The fit for the achondrites NWA 011, NWA 2976, and NWA 4587 (Fig. \ref{fig2}, third row, left panel) produced an acceptable fit quality of $\chi_{n}\leq 2.5$ for an accretion between $0$ and $0.6$ Ma and radii of $>20$ km. A minimum field can hardly be identified, if at all, by $\chi_{n}\leq 2.2$ that does not affect the accretion time interval, but changes the radius estimate to $>40$ km. Relatively high fit quality values are rooted in a maximum time difference of $\approx 0.7$ Ma at the depth of NWA 4587, very small uncertainties of Al-Mg ages of $0.33$, $0.05$, and $0.04$ MA for NWA 011, NWA 2976, and NWA 4587, respectively, and a necessity to fit each of these data points along with the respective Mn-Cr or Pb-Pb data at the same depth for each meteorite.

Finally, the fit for the grouplet comprising NWA 6704 and NWA 10132 was done with four data points. The shapes of the $\chi_{n}$ isolines resemble those obtained for the NWA 011 grouplet, as it was expected by the relative similarity of the data. Models produced an acceptable fit quality of $\chi_{n}\leq 1.8$ for $t_{0}\lesssim 0.9$ Ma and $R>20$ km, and a plateau with $\chi_{n}\leq 1.5$ for an accretion before $0.7$ Ma and a radius of $>30$ km. 

The right column of Fig. \ref{fig2} shows temperature curves at burial depths of the meteorites from which the data were obtained within the representative best-fit parent bodies. The latter are indicated by red dots in the respective fit quality panels. For CR1-3 (top right panel), two depths corresponding to two meteorites Renazzo (5.2 km) and GRO 95577 (7.8 km) were calculated within an object with a radius of $90$ km. The temperature curves fit the Mn-Cr data very well and the maximum values of $370$ K or $395$ K agree with the temperature ranges suggested from lab work and with the petrologic type. A larger burial depth and a longer heating phase of GRO 95577 are in agreement with its high alteration degree compared to Renazzo and corroborate, further, the positions of these meteorites on the CR mixing line as a result of their alteration extents.
From one data point available for Flensburg, only one burial depth within the parent body could be fitted (Fig. \ref{fig2}, second row, right panel). The temperature curve for the representative body with $R=20$ km crosses the Mn-Cr carbonate data point on its prograde branch in agreement with the data point reflecting the age of an early and prograde carbonate formation and the lack of carbonate forming fluids later on. A maximum temperature of $565$ K reached lies well within the interval of $470-670$ K suggested.\cite{Bischoff2021} 
In the case of NWA 011 grouplet (Fig. \ref{fig2}, third row, right panel), three meteorites with two data points each resulted in fits of three different burial depths within the parent bodies in the $\chi_{n}$ minimum field. Burial depths of $2.55$, $2.6$, and $3$ km within a body with $R=100$ km correlate with the decreasing Al-Mg age of the meteorites, reflecting stronger heating and slower cooling at higher depth. A burial depth difference of $\approx 50$ m for NWA 011 and NWA 4587 results from an Al-Mg age difference of only $0.18$ Ma. The maxima of the respective temperature curves of $1610-1630$ K are sufficient to produce basaltic melts consistent with the composition of the grouplet members and our model produced differentiated silicate material at the fit depth. However, depths of $2.3-3$ km within the sample parent body also implies an intrusive nature for these meteorites.
For the NWA 6704 grouplet, fits of three NWA 6704 and one NWA 10132 data points produce typically two different depths. Both fits are dominated by the Al-Mg age difference of $0.1$ Ma that resulted in a burial depth difference of $\approx 50$ m. A deeper location for NWA 10132 with an younger Al-Mg age is consistent with slightly higher temperature maximum and slower cooling. Similarly to the NWA 011 grouplet, maximum temperatures of $1645-1648$ K agree with silicate melting, while the differentiated silicate material obtained with our model at the burial depths agrees with the meteorite composition. Burial depths of $\approx 3.3$ km imply an intrusive origin also for this grouplet.

\begin{figure}
\begin{minipage}[ht]{8.5cm}
\setlength{\fboxsep}{0mm}
\includegraphics[trim = 20mm 5mm 20mm 0mm, width=8.5cm]{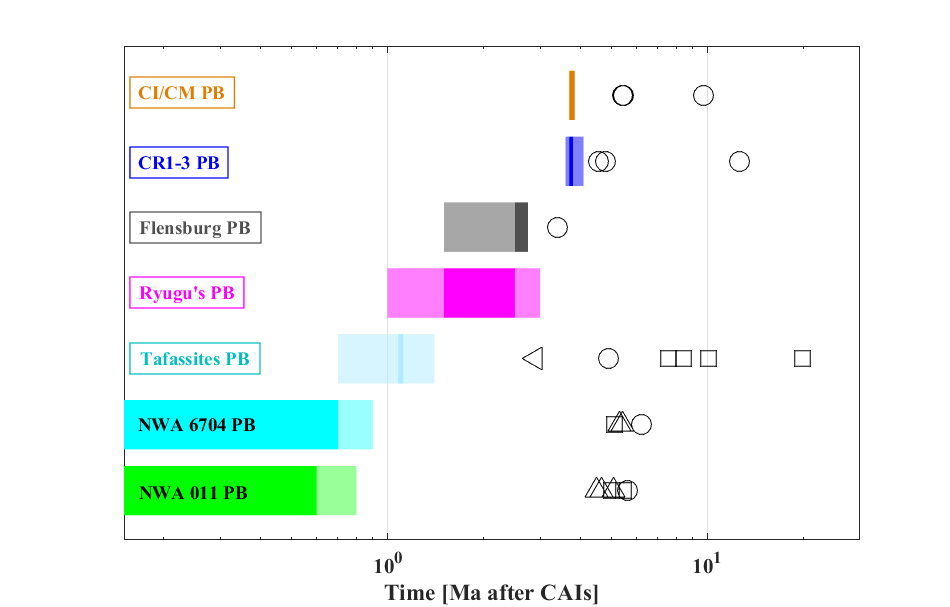}
\end{minipage}
\caption{Parent body accretion times (colored patches, dark for best-fits and transparent for acceptable fits) and meteorite metamorphic ages (data points) for the C reservoir modeled incl. Tafassites, Ryugu, and CI/CM PB.\cite{Ma2022,Neumann2021} Younger Mn-Cr carbonate ages (circles) of CR1-3 and CI/CM than for Flensburg result in moderately younger accretion age for the CR1-3 and CI/CM PBs. NWA 6704 (cyan) and NWA 011 (green) have older Pb-Pb whole-rock ages (squares) than Tafassites (light blue), resulting in earlier accretion than the Tafassite PB. The NWA 011 and NWA 6704 Al-Mg (triangles) and Mn-Cr (circles) data support an early PB formation. While Tafassites\cite{Ma2022} also include a very late, i.e., young, phosphate Pb-Pb age, their old iron-silicate Hf-W (triangle) and Mn-Cr ages cause a shift to an intermediate PB formation time between NWA 011 and NWA 6704 on one hand, and Flensburg, CR1-3, and CI/CM PBs on the other, while the young Pb-Pb age has its major effect in the large Tafassites PB size. 
All accretion times correlate inversely with the meteorite petrologic type and the degree of metamorphism or melting. See Table \ref{table1} for the thermo-chronological data.}
\label{fig3}
\end{figure}

\paragraph{Discussion}
Our calculations indicate an origin on different parent bodies for the meteorite groups considered, with a parent body accretion time of $t_{0}\approx 3.7$ Ma and size of $R \approx 90$ km for the CR1-3, $t_{0}\approx 2.5-2.75$ Ma for Flensburg, $t_{0}\lesssim 0.6$ Ma and $R>40$ km for NWA 011 grouplet, and $t_{0} \lesssim 0.7$ Ma and $R>30$ km  for NWA 011. This demonstrates a spread of formation times between $0$ Ma and $4$ Ma after the formation of the solar system in the accretion disk region of carbonaceous chondrites. Accretion times we derived recently for other water-rich parent bodies, such as Tafassites ($1.1_{-0.4}^{+0.3}$ Ma),\cite{Ma2022} Ryugu ($1-3$ Ma),\cite{Neumann2021} and a CM/CI asteroid ($\approx 3.8$ Ma)\cite{Neumann2021} support this conclusion. Our results confirm, further, that rocky parent bodies accreted early and sequentially\cite{Ma2022} from $<1$ Ma to $4$ Ma after CAIs, and witnessed different accretion mechanisms within a limited region of the outer solar system. The CR1-3 parent body complements the CI/CM parent body with a considerably smaller radius of $R\approx 15-25$ km but a similar accretion time of $3.75-3.8$ Ma and a sub-10 km sized parent body of the NEA Ryugu that accreted between $1$ Ma and $3$ Ma suggested recently.\cite{Neumann2021} Our results for the CR1-3 parent body and the Tafassites parent body model\cite{Ma2022} imply, further, that the meteorites classified initially as CR6-7 and re-classified as Tafassites do not belong to the same group as aqueously altered CR.
An earlier formation of the NWA 011 and NWA 6704 grouplets' parent bodies than that of the Tafassites is in an agreement with the former being differentiated and the latter only partially molten.
The parent body size estimates are less conclusive due to the lack of data but they fall into the range of typical sizes estimated for early solar system planetesimals.\cite{Delbo2017} Of note is a size contrast to the sub-10 km large parent body of Ryugu that was derived based on porosity modeling given lack of any chronological data in 2021.\cite{Neumann2021}

Two types of structures were obtained in the terms of metal-silicate differentiation. Parent bodies of CR1-3 and Flensburg are homogeneous with maximum temperatures remaining below the melting temperatures of metal or silicates. Internal heating by $^{26}$Al warmed large fractions of their volumes relatively uniformly, leaving a small volume of outer layer relatively cool, and supporting production of petrologic types 1-3 at different depths. These alteration conditions are similar to those of CI/CM PB,\cite{Neumann2021} while bearing similarity to a lesser extent to the original parent body of Ryugu that was suggested to be heated to a higher temperature that enabled a partial dehydration.\cite{Neumann2021,Sugita2019} In addition, peak central temperatures of up to $700$ K obtained for the favored Flensburg PB models also come close to those derived for Ryugu from a potential partial dehydration. While the CR1-3 carbonate ages are older than Flensburgs within the margin of error, an earlier accretion of the Flensburg asteroid than the CR1-3 PB derived confirms that it recorded the earliest fluid activity on a planetary object in the solar system.\cite{Bischoff2021} Initial Ryugu sample analysis indicates a close similarity to the CI material\cite{Yokoyama2022}, suggesting a potential origin on a common parent body. Ryugu samples originate from two different sites and have high porosities comparable with that derived by MASCOT that observed at yet another site. Thus, the asteroid is suggested to be homogeneous. However, an origin of high-porosity Ryugu material and low-porosity CI meteorites on a common parent body would require either a strong sorting of the rubble after parent body destruction, such that only high-porosity pieces would accrete to form Ryugu, but such a sorting appears unlikely. Or, it would require different impacts onto the parent body, of which one would be only shallow and eject enough material from a high-porosity surface layer to form Ryugu, and at least one other impact would eject CI material from deeper within where porosity has been reduced stronger.

An opposite differentiated structure type was produced due to partial melting in the interior of NWA 011 and NWA 6704 parent bodies. These structures are similar to those of primitive achondritic Acapulco-Lodran\cite{Neumann2018b} and Tafassites parent asteroids.\cite{Ma2022} They are, further, consistent with the accretion times calculated for them and reported for achondrite and primitive achondrites in the literature. Not much chondritic material with typical petrologicy types relative to the total parent body volume can be expected on these parent due to small burial depths for of overall $2.5-3.3$ km derived for the differentiated meteorites. By contrast, primitive achondritic and achondritic materials are abundant at these depths and further below. Any kind of chondritic material could be retained only in thin surface shells, while most of the parent body volume would be dominated by achondritic metal or silicate rocks. In particular, early accretion of the NWA 011 and NWA 6704 parent asteroids implies a similar magmatic evolution to Vesta that formed within $\lesssim 1$ Ma.\cite{Neumann2014} Bearing similarities to eucrites and classified originally as a non-cumulate eucrite,\cite{Afanasiev2000} NWA 011, and also NWA 6704 likely experienced similar petrogenesis. Given overall burial depths of $2.55-3.25$ km derived, both grouplets would be intrusive rocks and, in fact, NWA 011 has a cumulate-eucrite-like REE pattern,\cite{Mittlefehldt2007} while NWA 6704 spectrum showed a close similarity to the basaltic and cumulate eucrites.\cite{Lecorre2014} The asteroid (34698) 2001 OD22 suggested as parent body of NWA 6704\cite{Mcgraw2020} does not fulfill the size requirements we derived and, if related, should be a piece of a fractured parent asteroid.
A partially differentiated parent body with a radius of $\approx 200$ km has been proposed previously as parent body of CR1-3 and NWA 011 \cite{Bunch2005,Nehru2014,Irving2014}. Our results contradict that, both due to a smaller best-fit CR1-3 parent body size and a substantial difference of its accretion to that of NWA 011 and NWA 6704. 
A lack of sufficiently large asteroids with exposed silicates that would meet compositional constraints of CR-related meteorites suggests that their parent bodies were removed from the asteroid belt or have experienced mantle removal and their cores are represented by metallic asteroids. The parent body of the presumably metallic asteroid (16) Psyche should have a radius of $\approx 200$ km, since Psyche's mean radius is $\approx 110$ km, and would, therefore, lie in the best-fit fields for NWA 011 and NWA 6704. Regardless of parent body removal or loss of mantle, ejection of mantle portions by energetic collisions could have produced small basaltic meteorites present in the asteroid belt.

Accretion times of CR-related parent bodies including the result for Tafassites \cite{Ma2022} with other water-rich parent bodies from the C reservoir, such as that of Ryugu and of CI/CM groups, are compared in Fig. \ref{fig3}. The emerged accretion trend is correlated with the petrologic type of the meteorites and the degree of metamorphism or melting in agreement with the control of the heating rate by the amount of $^{26}$Al incorporated. The result of a temporally distributed accretion of CR-related parent bodies implies that accretion processes in the C reservoir started as early as in the NC reservoir and produced differentiated parent bodies with carbonaceous chondritic compositions, along with a spectrum of petrographic types, compositions, and parent body structures.
A confirmation of the Ryugu PB age and an addition of carbonaceous NEA formation times are expected from extensive Hayabusa2 and OSIRIS-REx sample analyses. Accretion in different parts of protoplanetary disks is observable by facilities such as ALMA, cosmochemical analyses of accretion processes are achievable using observations by VLT, while exoplanet search projects, such as PLATO and CHEOPS telescopes, can provide further clues via statistical information on the architecture of planetary systems.

\section{Funding:} WN acknowledges support by the Deutsche Forschungsgemeinschaft (DFG, project number 434933764). MT and WN acknowledge support by Klaus Tschira Foundation.

\section{Competing Interests:} None.

\section{Data and Materials Availability:} The data that support the plots within this paper and other findings of this study are available from the corresponding author upon reasonable request.

\section{Code Availability:} The numerical model used is described in detail in the Supplementary Information. The code is available only from the corresponding author upon reasonable request.

\section{Supplementary Materials:} Supplementary materials are available as a separate pdf file.



\begin{thebibliography}{9}
\bibitem{Sanborn2019} Sanborn, M. W., Wimpenny, J., Williams, C. D., Yamakawa, A., Amelin, Y., Irving, A. J., Yin, Q.-Z. Carbonaceous achondrites Northwest Africa 6704/6693: Milestones for early Solar System chronology and genealogy. Geochimica et Cosmochimica Acta 245, 577-596 (2019).
\bibitem{Huyskens2019} Huyskens, M. H., Sanborn, M. E., Yin, Q.-Z., Amelin, Y., Koefoed, P. Chronology of carbonaceous achondrites from the outer solar system. 50th Lunar and Planetary Science Conference 2019, abstract \#2736 (2019).
\bibitem{Clayton1993} Clayton, R. N. Oxygen isotopes in meteorites. Annual Review of Earth and Planetary Sciences 21, 115-149 (1993).
\bibitem{Baker2002} Baker, L., Franchi, I. A., Wright, I. P., Pillinger, C. T. The Oxygen isotopic composition of water from Tagish Lake: Its relationship to low-temperature phases and to other carbonaceous chondrites. Meteoritics and Planetary Science 37, 977-985 (2002).
\bibitem{Engrand2001} Engrand, C., Gounelle, M., Zolensky, M. E. In-SITU oxygen isotopic composition of Tagish Lake: An ungrouped type 2 carbonaceous chondrite. OAI identifier: oai:casi.ntrs.nasa.gov:20110011647, http://hdl.handle.net/2060/20110011647 (2001).
\bibitem{Bunch2005} Bunch, T. E., Irving, A. J., Larson, T. E., Longstaffe, F. J., Rumble, D. III, Wittke, J. H. "Primitive" and igneous achondrites related to the large and differentiated CR parent body. Lunar and Planetary Science XXXVI, abstract nr. 2308 (2005).
\bibitem{Nehru2014} Nehru, C. E., Boesenberg, J. S., Weisberg, M. K. Tafassasset and primitive achondrites: records of planetary differentiation. 77th Annual Meteoritical Society Meeting, abstract \#5382 (2014).
\bibitem{Irving2014} Irving, A. J., Kuehner, S. M., Tanaka, R., Rumble, D., Ziegler, K, Sanborn, M., Yin, Q. Collisional disruption of a layered, differentiated CR parent body containing metamorphic and igneous lithologies overlain by a chondrite veneer. Lunar and Planetary Science XLV, abstract nr. 2465 (2014).
\bibitem{Leshin2001} Leshin, L. A., Farquhar, J., Guan, Y., Pizzarello, S., Jackson, T. L., Thiemens, M. H. Oxygen isotopic anatomy of Tagish Lake: Relationship to primary and secondary minerals in CI and CM chondrites. Lunar and Planetary Science XXXII, abstract nr. 1843 (2001).
\bibitem{Yurimoto2008} Yurimoto, H., Krot, A. N., Choi, B.-G., Aleon, J., Kunihiro, T., Brearley, A. J. Oxygen isotopes of chondritic components. Reviews in Mineralogy and Geochemistry 68, 141-186 (2008).
\bibitem{Piralla2020} Piralla, M., Marrocchi, Y., Verdier-Paoletti, M. J., Vacher, L. G., Villeneuve, J., Piani, L., Bekaert, D. V., Gounelle, M. Primordial water and dust of the Solar System: Insights from in-situ oxygen measurements of CI chondrites. Geochimica et Cosmochimica Acta 269, 451-464 (2020).
\bibitem{Telus2019} Telus, M., Alexander, C. M., Hauri, E. H., Wang, J. Calcite and dolomite formation in the CM parent body: Insight from in situ C and O isotope analyses. Geochimica et Cosmochimica Acta 260, 275-291 (2019).
\bibitem{Hyde2013} Hyde, B. C., Tait, K. T., Nicklin, I., Grogory, D. A., Ali, A., Jabeen, I., Banerjee, N. R. Northwest Africa 7680: an ungrouped achondrite with affinities to primitive achondrite groups. 76th Annual Meteoritical Society Meeting, abstract \#5207 (2013).
\bibitem{Lindgren2017} Lindgren, P., Lee, M. R., Starkey, N. A., Franchi, I. A. Fluid evolution in CM carbonaceous chondrites tracked through the oxygen isotopic compositions of carbonates. Geochimica et Cosmochimica Acta 204, 240-251 (2017).
\bibitem{Fujiya2019} Fujiya, W., Hoppe, P., Ushikubo, T., Fukuda, K., Lindgren, P. Lee, M. R., Koike, M., Shirai, K., Sano, Y.  Migration of D-type asteroids from the outer Solar System inferred from carbonate in meteorites. Nature Astronomy 3, pages 910–915 (2019).
\bibitem{Benedix2003} Benedix, G. K., Leshin, L. A., Farquhar, J., Jackson, T., Thiemens, M. H. Carbonates in CM2 chondrites: Constraints on alteration conditions from oxygen isotopic compositions and petrographic observations. Geochimica et Cosmochimica Acta 67, 1577-1588 (2003).
\bibitem{Ma2022} Ma, N., Neumann, W., Neri, A., Schwarz, W. H., Ludwig, T., Trieloff, M., Klahr, H., Bouvier, A. Early formation of primitive achondrites in an outer region of the protoplanetary disc Geochemical Perspective Letters 23, 33-37 (2022).
\bibitem{Dauphas2016} Dauphas, N, Schauble, E. A. Mass Fractionation Laws, Mass-Independent Effects, and Isotopic Anomalies. Annual Review of Earth and Planetary Sciences, 44, 709-783 (2016).
\bibitem{Scott2018} Scott, E. R. D., Krot, A. N., Sanders, I. S. Isotopic Dichotomy among Meteorites and Its Bearing on the Protoplanetary Disk. The Astrophysical Journal, 854, 164 (2018).
\bibitem{Burkhardt2019} Burkhardt, C., Dauphas, N., Hans, U., Bourdon, B., Kleine, T. Elemental and isotopic variability in solar system materials by mixing and processing of primordial disk reservoirs. Geochimica et Cosmochimica Acta, 261, 145-170 (2019).
\bibitem{Tyra2007} Tyra, M. A., Farquhar, J., Wing, B. A., Benedix, G. K., Jull, A. J. T., Jackson, T., Thiemens, M. H. Terrestrial alteration of carbonate in a suite of Antarctic CM chondrites: Evidence from oxygen and carbon isotopes. Geochimica et Cosmochimica Acta 71, 782-795 (2007).
\bibitem{Tyra2012} Tyra, M. A., Farquhar, J., Guan, X., Leshin, L. A. An oxygen isotope dichotomy in CM2 chondritic carbonates - A SIMS approach. Geochimica et Cosmochimica Acta 77, 383-395 (2012).
\bibitem{Tyra2013} Tyra, M. A. Using oxygen and carbon stable isotopes, 53Mn-53Cr isotope systematics, and petrology to constrain the history of carbonates and water in the CR and CM chondrite parent bodies. University of New Mexico Albuquerque, New Mexico, UNM Digital Repository, https://digitalrepository.unm.edu (2013).
\bibitem{Tyra2016} Tyra, M., Brearley, A., Guan, Y. Episodic carbonate precipitation in the CM chondrite ALH. Geochimica et Cosmochimica Acta 175, 195-207 (2016).
\bibitem{Schrader2011} Schrader, D. L., Franchi, I. A., Connoly, H. C. Jr., Greenwood, R. C., Lauretta, D. S., Gibson, J. M. The formation and alteration of the Renazzo-like carbonaceous chondrites I: Implications of bulk-oxygen isotopic composition. Geochimica et Cosmochimica Acta 75, 308-325 (2011).
\bibitem{Clayton1976} Clayton, R. N., Onuma, N., Mayeda, T. K. A classification of meteorites based on oxygen isotopes. Earth and Planetary Science Letters 30, 10-18 (1976).
\bibitem{Clayton1999} Clayton, R. N., Mayeda, T. K. Oxygen isotope studies of carbonaceous chondrites. Geochimica et Cosmochimica Acta 63, 2089-2104 (1999).
\bibitem{Rowe1994} Rowe, M. W., Clayton, R. N., Mayeda, T. K. Oxygen isotopes in separated components of CI and CM meteorites. Geochimica et Cosmochimica Acta 58, 5341-5347 (1994).
\bibitem{Bischoff2021} Bischoff, A., Alexander, C. M. O'D., Barrat, J.-A., et al. The old, unique C1 chondrite Flensburg – Insight into the first processes of aqueous alteration, brecciation, and the diversity of water-bearing parent bodies and  lithologies. Geochimica et Cosmochimica Acta 293, 142-186 https://doi.org/10.1016/j.gca.2020.10.014 (2021).
\bibitem{Neumann2021} Neumann, W., Grott, M., Trieloff, M., Jaumann, R., Biele, J., Hamm, M., Kührt, E. Microporosity and parent body of the rubble-pile NEA (162173) Ryugu. Icarus 358, 114166, https://doi.org/10.1016/j.icarus.2020.114166, (2021).
\bibitem{Weiss2013} Weiss, B. P., Elkins-Tanton, L. T. Differentiated planetesimals and the parent bodies of chondrites. Annual Review of Earth and Planetary Sciences 41, 529-260 (2013).
\bibitem{Bouvier2011} Bouvier, A., Spivak-Birndorf, L. J., Brennecka, G. A., Wadhwa, M. New constraints on early Solar System chronology from Al-Mg and U-Pb isotope systematics in the unique basaltic achondrite Northwest Africa 2976. Geochimica et Cosmochimica Acta 75, 5310-5323 (2011).
\bibitem{Amelin2019} Amelin, Y., Koefoed, P., Iizuka, T., Fernandes, V. A., Huyskens, M. H., Yin, Q.-Z., Irving, A. J. U-Pb, Rb-Sr and Ar-Ar sysematics of the ungrouped achondrites Northwest Africa 6704 and Northwest Africa 6693. Geochimica et Cosmochimica Acta 245, 628-242 (2019).
\bibitem{Amelin2019a} Amelin, Y., Rydeblad, E., Koefoed, P. Krestianinov, E., Huyskens, M. H., Yin, Q.-Z. Pb-isotopic and initial Sr ages of the achondrite NWA 4587. 50th Lunar and Planetary Science Conference, abstract \#2261 (2019).
\bibitem{Steffens2019} Steffens, S. Mn-Cr Datierungen meteoritischer Lithologien. Bachelor thesis, Heidelberg University (2019).
\bibitem{Connelly2012} Connelly, J. N., Bizarro, M., Krot, A. N., Nordlund, A., Wielandt, D., Ivanova, M. A. The Absolute Chronology and Thermal Processing of Solids in the Solar Protoplanetary Disk. Science 338, 651-655 (2012).
\bibitem{Jilly2015} Jilly-Rehak, C. E., Huss, G. R., Nagashima, K. Oxygen isotopes in secondary minerals in CR chondrites: comparing components of different petrologic type. XLVI Lunar and Planetary Science Conference, \#1662 (2015).
\bibitem{Jilly2017} Jilly-Rehak, C. E., Huss, G. R., Nagashima, K. 53Mn-53Cr radiometric dating of secondary carbonates in CR chondrites: Timescales for parent body aqueous alteration. Geochimica et Cosmochimica Acta 201, 224-244 (2017).
\bibitem{Jilly2018} Jilly-Rehak, C. E., Huss, G. R., Nagashima, K., Schrader, D. L. Low-temperature aqueous alteration on the CR chondrite parent body: implications from in situ oxygen-isotope analyses. Geochimica et Consmochimica Acta 222, 230-252 (2018).
\bibitem{Goepel1994} Göpel, C., Manhes, G., Allegre, C. U-Pb systematics of phosphates from equilibrated ordinary chondrites. Earth and Planetary Science Letters 121, 153-171 (1994).
\bibitem{Dodson1973} Dodson, M. H. Closure temperature in cooling geochronological and petrological systems. Contributions to Mineralogy and Petrology 40, 259-274 (1973).
\bibitem{Weisberg1993} Weisberg, M. K., Prinz, M., Clayton, R. N., Mayeda, T. K. The CR (Renazzo-type) carbonaceous chondrite group and its implications. Geochimica et Cosmochimica Acta 57, 1567–1586 (1993).
\bibitem{Weisberg1995} Weisberg, M., Prinz, M., Clayton, R., Mayeda, T., Grady, M., Pillinger, C. The CR chondrite clan. Proceedings of the NIPR Symposium on Antarctic Meteorite 8, 11 (1995).
\bibitem{Bottke2017} Bottke, W. and Morbidelli, A. Planetesimals: Early Differentiation and Consequences for Planets Ch. Using the Main Asteroid Belt to Constrain Planetesimal and Planet Formation (Cambridge University Press, 2017).
\bibitem{Henke2012} Henke, S., Gail, H.-P., Trieloff, M., Schwarz, W. H., Kleine, T. Thermal evolution and sintering of chondritic planetesimals. Astronomy and Astrophysics 537, A45 (2012).
\bibitem{Gail2019} Gail, H.-P., Trieloff, M. Thermal history modelling of the L chondrite parent body. Astronomy and Astrophysics 628, A77 (2019).
\bibitem{Neumann2018b} Neumann, W., Henke, S., Breuer, D., Gail, H.-P., Schwarz, W. H., Trieloff, M., Hopp, J., Spohn, T. Modeling the evolution of the parent body of acapulcoites and lodranites: A case study for partially differentiated asteroids. Icarus 311, 146-169 (2018).
\bibitem{Budde2018} Budde, G., Kruijer, T. S., Kleine, T. Hf-W chronology of CR chondrites: Implications for the timescales of chondrule formation and the distribution of 26Al in the solar nebula. Geochimica et Cosmochimica Acta 222, 284-304 (2018).
\bibitem{Sugita2019} Sugita, S., Honda, R., Morota, T., Kameda, S., Sawada, H., Tatsumi, E., Yamada, M., Honda, C., Yokota, Y., Kouyama, T., Sakatani, N., Ogawa, K., Suzuki, H., Okada, T., Namiki, N., Tanaka, S., Iijima, Y., Yoshioka, K., Hayakawa, M. The geomorphology, color, and thermal properties of Ryugu: Implications for parent-body processes. Science 364, eaaw0422 (2019).
\bibitem{Delbo2017} Delbo, M., Walsh, K., Bolin, B., Avdellidou, C., Morbidelli, A. Identification of a primordial asteroid family constrains the original planetesimal population. Science 357, 1026–1029 (2017).
\bibitem{Mittlefehldt2007} Mittlefehldt, D. W. 1.11 - Achondrites. Treatise on Geochemistry, Editor(s): Heinrich D. Holland, Karl K. Turekian, 1-40, ISBN 9780080437514, https://doi.org/10.1016/B0-08-043751-6/01064-1, (2007).
\bibitem{Afanasiev2000} Afanasiev, S. V., Ivanova, M. A., Korotchantseva, E. V., Kononkova, N. N.,  Nazarov,  M.  A. Dhofar  007  and  Northwest Africa  011: Two new eucrites of different types (abstract). Meteoritics \& Planetary Science 35, A19 (2000).
\bibitem{Neumann2014} Neumann, W., Breuer, D., Spohn, T. Differentiation of Vesta: Implications for a shallow magma ocean. Earth and Planetary Science Letters 395, 267-280 (2014).
\bibitem{Lecorre2014} Le Corre, L., Reddy, V., Cloutis, E. A., Mann, P., Buchanan, P. C., Gabelica, Z., Hupe, G., Gaffey, M. J. Identifying parent asteroid of ungrouped achondrites Northwest Africa 6704: lessons from Dawn at Vesta. 45th Lunar and Planetary Science Conference 2014, abstract \#1311 (2014).
\bibitem{Mcgraw2020} McGraw, A. M., Reddy, V., Izawa, M. R. M., Sanchez, J. A., Le Corre, L., Cloutis, E. A., Applin, D. M., Pearson, N. Mineralogical criteria for the parent asteroid of the "carbonaceous" achondrite NWA 6704. The Astronomical Journal 159, 107 (11pp) (2020).

\bibitem{Yokoyama2022} Yokoyama, T., 149 co-authors. Samples returned from the asteroid Ryugu are similar to Ivuna-type carbonaceous meteorites. Science, DOI: 10.1126/science.abn7850 (2022).



\end{thebibliography}
\end{document}